\documentclass[epj,final,nopacs]{svjour}
\usepackage{macros,cite}
\usepackage{amssymb}
\usepackage{epsfig}
\begin{document}
%
%%%%%%%%%%%%%%%%%%%%%%%%%%%%%%%%%%%%%%%%%%%%%%%%%%%%%%%%%%%%%%%%%%%%%%%%%%%%%
%%%%% Title Page Compatible with Springer Verlag's svjour Style File %%%%%%%%
%%%%%%%%%%%%%%%%%%%%%%%%%%%%%%%%%%%%%%%%%%%%%%%%%%%%%%%%%%%%%%%%%%%%%%%%%%%%%
%
%%%%% redefine front page header of "svjour.cls" to include 
%%%%% preprint number and logo  
%
\def\makeheadbox{{
\vspace{-0.25cm}
\begin{flushleft}
\vbox{
\epsfxsize=2.5 true cm
\epsfbox{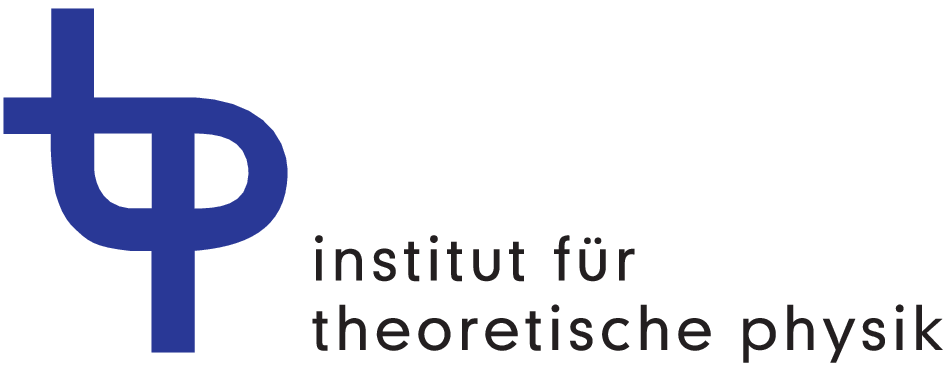}}
\end{flushleft}
\begin{flushright}
\vspace{-0.75cm}
MS-TP-03-23
\end{flushright}
\hss}}
%
%%%%% from here on: standard syntax (apart from an adjusting "\vspace") 
%
\title{
\vspace{-0.75cm}\\
A non-perturbative computation of the
B-meson decay constant and the b-quark mass in HQET%
\thanks{Invited talk at the 
{\it International Europhysics Conference on High Energy Physics}, 
EPS-HEP2003, July 17 -- 23, 2003, in Aachen, Germany.}
}
\author{
Jochen Heitger 
(ALPHA Collaboration)
}
\institute{
Universit\"at M\"unster, Institut f\"ur Theoretische Physik,
Wilhelm-Klemm-Str.~9, D-48149 M\"unster, Germany
}
\date{}
\abstract{
A lattice computation of the B-meson decay constant and the mass of the 
b-quark to leading order in the heavy quark effective theory is presented. 
The involved renormalization problems are solved non-perturbatively, and 
the continuum limit is taken.
In the quenched approximation the results reported here already offer an 
interesting numerical precision, which will be further improved in the 
near future.
}
\authorrunning{
Jochen Heitger
}
\titlerunning{
A non-perturbative computation of the
B-meson decay constant and the b-quark mass in HQET
}
\maketitle

\section{Introduction}
\label{Sec_intro}
Not least by the influence of the phenomenologically very interesting 
programme of current experiments to investigate CP-violation in the 
B-system 
% \cite{Back:2003ty,Yamauchi:2003rw,Zoccoli:2003ih},
\cite{BphysExp:2003}, 
the study of B-meson physics has become a vivid area of research.
To interpret the experimental observations within (or beyond) the standard 
model, matrix elements between low-energy hadron states must be known.
But since these QCD matrix elements live in the strongly coupled 
sector of the theory, they naturally call for a genuinely non-perturbative, 
`ab initio' approach for their determination:
the lattice formulation of QCD, which enables a numerical computation of
its low-energy properties through Monte Carlo evaluation of the underlying
Euclidean path integral.

Lattice QCD calculations with b-quarks can valuably contribute to precision 
CKM-physics by (over-)con\-strain\-ing the unitarity triangle and help to 
obtain other phenomenologically relevant predictions.
Examples for experimentally inaccessible key parameters that are important
here are the B-meson decay constant and the mass of the b-quark, which are 
subject of the present report.
In studying B-physics on the lattice, however, we face some particular 
problems.
A first one already originates in the b-quark itself, the mass of which is 
much larger than the inverse lattice spacings, $1/a$, affordable in 
simulations on present-day computers even in the quenched approximation
($1/\mb\approx 1/(4\,\GeV)< a \approx 0.07\,\Fm$): 
huge discretization errors would render a realistic treatment of B-systems 
with a propagating b-quark on the lattice impossible.

This motivates to recourse to \emph{effective theories}.
Theoretically most attractive is the heavy quark effective theory (HQET) 
whose Lagrangian in lattice formulation
\[
\LHQET= 
\heavyb D_0\heavy+{\T \frac{1}{m}}\,\heavyb\left\{
-{\T \frac{\ckin}{2}}\,{\bf D}^2-\csig({\bf B}\sigma)\right\}\heavy
\,+\,\ldots
\]
is, to first order in the inverse heavy quark mass $1/m$, formally identical
to the continuum one. 
As a similar expansion holds for the matrix elements in question, lattice
HQET constitutes a systematic expansion in terms of $1/\mb$ for B-mesons
at rest \cite{stat:eichhill1} that also has a continuum limit order by
order in the $1/m$--expansion.

Mainly because of two reasons, it has not received much attention in the 
past though:
\begin{enumerate}
\item The rapid growth of statistical errors as the time separation of 
      correlation function is made large.
      This unwanted feature is already encountered in the lowest-order
      effective theory (static approximation) and limits a reliable
      extraction of masses and matrix elements.
\item The number of parameters in the effective theory does not only 
      increase with the order of the expansion, but they have also 
      to be determined non-perturbatively, since otherwise 
      --- as a consequence of the mixings among operators of different 
      dimensions allowed in the cutoff theory (e.g.~of 
      $\frac{1}{2m}\heavyb{\bf D}^2\heavy$ with $\heavyb D_0\heavy$) --- 
      one is always left with a perturbative remainder that diverges as 
      $a\rightarrow 0$.
      Hence, these power-law divergences cause the continuum limit not to 
      exist unless the theory is renormalized 
      \emph{non-perturbatively} \cite{Maiani:1992az}. 
\end{enumerate}

Here I summarize recent progress in both directions, which reflects in two 
concrete applications in the combined static and quenched approximation. 
These are a determination of the $\Bs$-meson decay constant, where a 
correction due to the finite mass of the b-quark is estimated by 
interpolating between the static result and 
$\Fds$ \cite{fbstat:pap1,lat03:fbstat}, and a fully non-perturbative 
computation of the b-quark's mass based on the idea of a 
\emph{non-perturbative matching of HQET and QCD in finite volume} as 
proposed in \Refs{lat01:mbstat,mbstat:pap1}.

\section{The $\Bs$-meson decay constant}
\label{Sec_fB}
On our way towards a precision computation of $\Fbs$ in quenched 
QCD \cite{fbstat:pap1,lat03:fbstat} we employ a two-step strategy.
First, the decay constant is calculated in lowest order of HQET, and then it 
is combined with available results for the pseudoscalar decay constant 
$\Fps(\mps)$ in QCD around the charm quark mass region by interpolation 
in $1/\mps$. 

\begin{figure}[htb]
\centering
\vspace{-2.0cm}
\epsfig{file=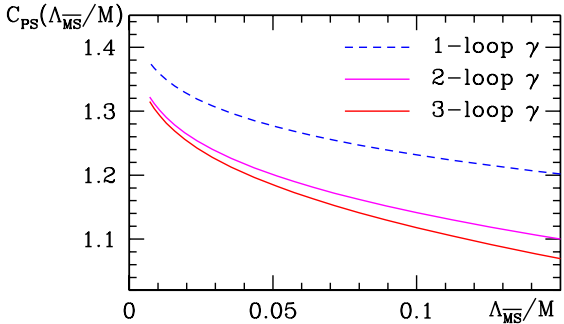,width=8.0cm}
\vspace{-2.0cm}
\caption{
Conversion factor to the matching scheme that translates a RGI matrix 
element of $\Astat$ at infinite mass to the one at finite mass.
Its uncertainty is estimated to be smaller than 2\% (half of the difference
between the 2-- and 3--loop results).
}\label{fig:Cps}
\vspace{-1.0cm}
\end{figure}

The pseudoscalar decay constant at finite mass is related to the 
\emph{renormalization group invariant (RGI)} matrix element of the static 
axial current,
\be
\PhiRGI\equiv\ZRGI\ketbra{\,{\rm PS}\,}{\,\Astat\,}{\,0\,}\,,
\label{me_RGI}
\ee
where $\Astat=\bpsis\gamma_0\gfv\psib^{\rm stat}$ in case of 
${\rm PS}=\Bs$, through:
\be
\Fps\sqrt{\mps}=
\Cps\left(M/\lMSbar\right)\times\PhiRGI\,+\,\Or\left(1/M\right)\,. 
\label{me_QCD}
\ee
Here, $M$ denotes the RGI mass of the heavy quark and $\lMSbar$ the QCD
$\Lambda$--parameter in the $\MS$ scheme.
The renormalization factor $\ZRGI$, turning any bare matrix element of 
$\Astat$ into the RGI one, has been non-perturbatively determined 
in \cite{zastat:pap3}.
$\Cps$ accounts for the fact that in order to extract predictions for QCD 
from results computed in the effective theory, its matrix elements are to be 
related to those in QCD at finite quark mass values.
In this sense $\Cps$ translates to the 
\emph{`matching scheme'} \cite{lat02:rainer,zastat:pap3}, which is 
defined by 
the condition that matrix elements in the (static) effective theory, 
renormalized in this scheme and at scale $\mu=m$, equal those in QCD up to 
$1/m$--corrections.
In leading order it is given via the large-mass asymptotics
\be
\PhiRGI=
\lim_{M\to\infty} 
\left[\,\ln(M/\lMSbar)\,\right]^{\gamma_0/(2b_0)}\Fps\sqrt{\mps}\,, 
\label{PhiRGI}
\ee
and thanks to the recent 3--loop computation of the anomalous dimension of
the static axial current \cite{ChetGrozin}, the function 
$\Cps(M/\lMSbar)=\Fps\sqrt{\mps}/\PhiRGI$ is now known 
perturbatively up to and including $\gbar^4(\mbar)$--corrections.
A numerical evaluation as explained in \cite{zastat:pap3} is shown 
in \Fig{fig:Cps}, where one can also infer that the remaining perturbative 
uncertainty has become very small.
%
%%%%%% figure: C_PS(Lambda_MSbar/M)
%
\subsection{RGI matrix element in the static theory}
As mentioned before, heavy-light correlation functions on the lattice, from 
which B-physics matrix elements such as the B-meson decay constant in 
question are obtained at large Euclidean time, are affected by large 
statistical errors in the static approximation.
Their noise-to-signal ratio grows exponentially with the time separation,
and in particular for the Eichten-Hill action \cite{stat:eichhill1},
\be
\Sstat=a^4\sum_x\heavyb(x)D_0\heavy(x)\,,
\label{action_eh}
\ee
this ratio roughly behaves as 
$\exp\{x_0(\Estat-\mpi)\}$ \cite{stat:hashi} with $\Estat$ the bare ground 
state energy of a B-meson in the static theory, diverging linearly in the 
continuum limit.

\begin{figure}[htb]
\centering
\vspace{-0.25cm}
\epsfig{file=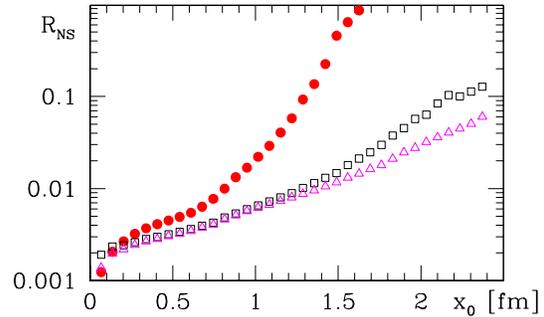,width=7.0cm}
\vspace{-0.125cm}
\caption{
Noise-to-signal ratio of a B-meson correlation function in static
approximation at $a\approx 0.08\,\Fm$ for various 
actions \protect\cite{fbstat:pap1}.
Bullets refer to the original Eichten-Hill action, while triangles 
correspond to the alternative discretization with HYP-links.
}\label{fig:rns}
\vspace{-0.375cm}
\end{figure}

To overcome this difficulty, we introduced in \Ref{fbstat:pap1} a few
alternative discretizations of the static theory that retain the $\Or(a)$
improvement properties of the action (\ref{action_eh}) but lead to a 
substantial reduction of the statistical fluctuations.
These new static quark actions rely on changes of the parallel transporters 
$U(x,\mu)$ in the covariant derivative
$D_0\heavy(x)=a^{-1}
[\,\heavy(x)-U^\dagger(x-a\hat{0},0)\heavy(x-a\hat{0})\,]$ of the form 
$U(x,0)\rightarrow W(x,0)$, where now $W(x,0)$ is a function of the gauge 
fields in the immediate neighbourhood of $x,x+a\hat{0}$.
Its best version employs `HYP-smearing' that takes for $W(x,0)$ the 
so-called HYP-link, which is a function of the gauge links located within a 
hypercube \cite{HYP:HK01}.
Comparing the noise-to-signal ratios in \Fig{fig:rns}, one can see that 
around $x_0\approx 1.5\,\Fm$ more than an order of magnitude can be gained in 
this case w.r.t.~to the Eichten-Hill action and, in addition, the 
statistical errors grow only slowly as $x_0$ is increased.
%
%%%%%% figure: noise-to-signal ratio of various static actions
%
Even more importantly, we observed \cite{fbstat:pap1} 
(see also \cite{lat03:statprec}) quite the same, small lattice artifacts 
with the new discretizations.

In our computational setup to determine the bare matrix element
$\ketbra{\Bs}{\Astat}{0}$ entering \eq{me_RGI} we use the Schr\"o\-dinger 
functional (SF) formulation of QCD with non-per\-tur\-batively $\Or(a)$ 
improved Wilson actions in the gauge and light (i.e.~relativistic) quark 
sectors.
For technical details and the exact definitions of the correlation functions 
we refer to \Refs{zastat:pap3,fbstat:pap1}.
Here we only record that \cite{msbar:pap2} 
\be
\ketbra{\,\Bs\,}{\,\Astat\,}{\,0\,}\propto
\frac{\fastat(x_0)}{\sqrt{\fone}}\,\Exp^{\,(x_0-T/2)\Estat(x_0)}\,,
\label{me_CFs}
\ee
modulo volume factors, where $\fastat$ is a proper SF correlation function 
of the ($\Or(a)$ improved) static axial current with the quantum numbers of 
a B-meson and $\fone$ is a corresponding boundary-to-boundary correlator, 
which serves to cancel the renormalization factors of the boundary quark
fields. 
Moreover, we implement wave functions at the boundaries of the SF-cylinder 
to construct an interpolating B-meson field that suppresses unwanted 
contaminations from excited B-meson states to the correlators. 

\begin{figure}[htb]
\centering
\vspace{-2.25cm}
\epsfig{file=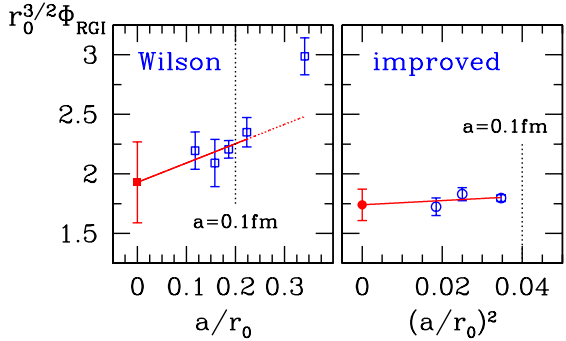,width=8.0cm}
\vspace{-1.75cm}
\caption{
RGI matrix element of $\Astat$ \protect\cite{zastat:pap3}, based on 
unimproved bare data from \protect\cite{stat:fnal2,stat:DMcN94} (left) and 
on simulations of the $\Or(a)$ improved theory with the new 
discretization \protect\cite{fbstat:pap1,lat03:statprec} (right).
A continuum extrapolation of the latter yields 
$r_0^{3/2}\PhiRGI=1.74(13)$.
}\label{fig:PhiRGI}
\vspace{-0.375cm}
\end{figure}

So far, the bare matrix element (\ref{me_CFs}) has been calculated for three
lattice spacings ($a$ $\approx$ $0.1\,\Fm$, $0.08\,\Fm$ and $0.07\,\Fm$), 
and the regularization dependent part of the factor $\ZRGI$, which according 
to (\ref{me_RGI}) must be attached to get the RGI matrix element, has been 
computed for the new actions as it was done for the Eichten-Hill action 
in \Ref{zastat:pap3}.
The continuum extrapolation quadratic in the lattice spacing of our results 
stemming from the static action with HYP-links is displayed in the right 
part of \Fig{fig:PhiRGI}.
%
%%%%%% figure: r_0^{3/2}*Phi_RGI
%
To illustrate the gain in precision and control of the systematic errors,
we confront our $\Or(a)$ improved results with an analysis of older 
unimproved Wilson data for the bare matrix element, reproduced 
from \cite{zastat:pap3} in the left part of the figure.
\subsection{Extrapolation in the heavy quark mass}
\begin{figure}[htb]
\centering
\vspace{-0.875cm}
\epsfig{file=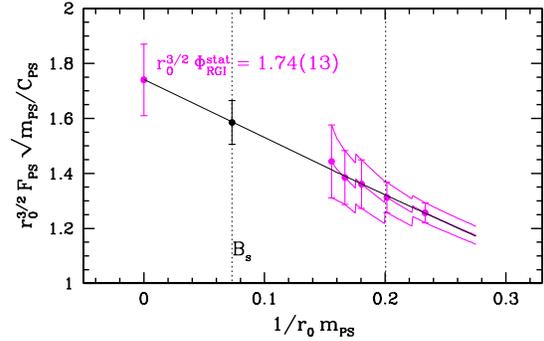,width=8.0cm}
\vspace{-0.525cm}
\caption{
Interpolation in the inverse heavy-strange meson mass, $\mps$, between the 
RGI matrix element of the static axial current and relativistic data around 
the charm quark mass \protect\cite{lat03:fbstat}.  
}\label{fig:interpol}
\vspace{-0.25cm}
\end{figure}

To finally arrive at a value for $\Fbs$, we combine $\PhiRGI$, referring to 
the static limit and thus being independent of the heavy quark mass, with 
numbers of $\Fps$ in the continuum limit at finite values of the quark mass, 
which have been collected in the context of the (quenched) computation of 
$\Fds$ of \cite{fds:JR03,lat03:fds}.
In incorporating the mass dependence (\ref{me_QCD}) predicted by HQET, we
are then led to extrapolate $r_0^{3/2}\Fps\sqrt{\mps}/\Cps(M/\lMSbar)$ 
from the charm region to the static estimate $r_0^{3/2}\PhiRGI$ by a linear 
fit in $1/(r_0\mps)$.
This interpolation is shown in \Fig{fig:interpol}, where the zigzag error
bands around the relativistic data indicate a small systematic effect that 
is due to the mass dependence of the discretization errors in the decay
constant near the charm quark mass as discussed 
in \Refs{lat03:fbstat,lat03:fds}.
While an extrapolation in $1/(r_0\mps)$ from the charm region without the 
constraint through the static approximation would look similar, it is 
obvious that the \emph{interpolation} is much safer, since extrapolating to 
the (quite distant) $\Bs$-meson scale depends significantly on the 
functional form assumed.

Using $\mBs=5.4\,\GeV$, $r_0=0.5\,\Fm$ and the numerical perturbative value 
of the matching factor $\Cps(\Mb/\lMSbar)$ translating to finite b-quark 
mass, we find from the interpolation to $1/(r_0\mBs)$ 
in \Fig{fig:interpol} as our present result \cite{lat03:fbstat}
\be
r_0\Fbs=0.52(3) \,\,\rightarrow\,\,
\Fbs=206(10)\,\MeV\,.
\ee
This number includes all errors except for the quenched approximation;
the (unavoidable) scale ambiguity introduced by it can be estimated to be
about $12\%$.
%
%%%%%% figure: interpolation of r_0^{3/2}*Phi_RGI in 1/(r_0*m_PS)
%

\section{The b-quark's mass}
\label{Sec_Mb}
The second of the aforementioned problems that so far hampered the use of 
HQET on the lattice is the occurrence of power-law divergences in the 
lattice spacing. 
It already shows up in the static approximation and thereby affects, for 
instance, the computation of the mass of the b-quark in leading order of 
HQET.
In this case the kinetic and the mass terms in the static action mix under 
renormalization and give rise to a local mass counterterm 
$\delta m\propto 1/a$, the self-energy of the static quark, which implies a 
linearly divergent truncation error if one relies on an only perturbative 
subtraction of this divergence.
Therefore, past lattice computations in the framework of HQET could not 
reach the continuum limit \cite{lat01:ryan,reviews:ichep02}.

A strategy for a solution to this longstanding problem was developed 
in \Ref{mbstat:pap1}, which now offers the possibility to perform clean, 
non-perturbative calculations in HQET.
It basically consists of three parts that I want to briefly describe in the
following by sketching a (still ongoing) computation of the b-quark's mass 
as example \cite{mbstat:pap1,mbstat:pap2}:
\begin{enumerate}
\item Renormalization of the effective theory amounts to relate the 
      parameters of the HQET Lagrangian to those of QCD, a step usually 
      called \emph{matching}.
      In order to realize the matching in a \emph{non-perturbative} way, one
      imposes matching conditions of the form 
      $\PhiHQET(L_0,M)=\PhiQCD(L_0,M)$ in a \emph{physically small volume}
      of linear extent $L_0=\Or(0.2\,\Fm)$, where $\PhiHQET$ and $\PhiQCD$
      are suitably chosen observables in HQET and QCD to be calculated with 
      the aid of numerical simulations.
      The finiteness of the matching volume ensures that lattice 
      resolutions satisfying $a\mb\ll 1$ are possible and the b-quark can be 
      treated as standard relativistic fermion, while at the same time the 
      energy scale $1/L_0=\Or(1\,\GeV)$ is still significantly below $\mb$ 
      and HQET applies quantitatively.
      In determining the parameters of the effective theory from those of 
      QCD via such a non-perturbative matching in finite volume, the 
      predictive power of QCD is transfered to HQET.
      Of course, owing to the very construction of the effective theory, it
      is clear that these matching conditions must also carry a dependence 
      on the heavy quark mass, which is most conveniently identified with the 
      (scheme and scale independent) \emph{RGI} mass, 
      $M$ (see e.g.~\cite{msbar:pap1}).
        
      In the concrete case of the b-quark mass computation, definite
      choices for the quantities $\Phi$ have to be made to formulate a 
      sensible matching condition between the quark mass in the full, 
      relativistic theory (QCD) and HQET. 
      Those are $\Gamma(L,\Mb)$, denoting the energy of a state with the
      quantum numbers of a B-meson but defined in a \emph{small volume} 
      of extent $L$, and $\gamstat(L)$ as its counterpart in 
      (leading order of) the effective theory.
      As detailed in \cite{mbstat:pap1,QCDvsHQET:pap1}, both can be 
      expressed as logarithmic derivatives of appropriate finite-volume, 
      heavy- and static-light correlation functions, respectively, and 
      numerically evaluated with high precision.
\item Next we need to establish a connection to a physical situation, where 
      observables of the infinite-volume theory such as masses or matrix 
      elements are accessible at the end.
      The accompanying gap between the small volume with its fine lattice 
      resolution, where the matching of HQET and QCD is done, on the one 
      side and larger lattice spacings (and also larger volumes) on the 
      other is bridged by a recursive finite-size scaling procedure inspired 
      by \cite{alpha:sigma}:
      the volume to compute the quantity $\gamstat(L)$ in HQET is 
      iteratively enlarged until one reaches a volume of linear extent 
      $L=\Or(1\,\Fm)$ so that, at the same resolutions $a/L$ (i.e.~at the 
      same bare parameters) met there, \emph{large volumes} with 
      $L\approx 2\,\Fm$ --- to accommodate physical observables in the 
      infinite-volume theory --- eventually become affordable.
      Also note that, apart from terms of $\Or(1/(L_0\Mb)^{n+1})$ if 
      considering HQET up to order $n$, any dependence on the unphysical 
      small-volume physics is gone now.
\item Finally, a physical, dimensionful input is still missing.
      In the case at hand this means to link the energy $\gamstat$, which 
      turns into the B-meson's static binding energy, $\Estat$, as the 
      volume grows, to the mass of the B-meson as the physical observable in
      large volume whose numerical value is taken from experiment.
\end{enumerate}
To join the foregoing three steps, we have to recall that energies in the
effective theory differ from the corresponding ones in QCD by a linearly
divergent mass shift $\mhbare$, which has its origin in the mixing of 
$\heavyb D_0\heavy$ with the lower-dimensional operator $\heavyb\heavy$
under renormalization --- the central problem we started from.
As a consequence of its universality (i.e.~its independence of the state), 
however, $\mhbare$ obeys at any fixed lattice spacing 
\bea
\mB 
& = & 
\Estat+\mhbare\,, 
\label{eq_mB}\\
\Gamma(L,\Mb)
& = &
\gamstat(L)+\mhbare\,.
\label{eq_Gam}
\eea
Imposing \eq{eq_Gam} for $L=L_0$ as the non-perturbative matching condition 
in small volume implicitly determines the parameter $\mhbare$ and may hence 
be exploited to replace it in \eq{eq_mB}.
Then, after adding and subtracting a term $\gamstat(L_2)$ (where 
$L_2=2^2L_0\approx 1\,\Fm$ with lattice spacings commonly used in 
large-volume simulations), the resulting equation may be cast into the 
basic formula 
\bea
\mB
& \,=\, &
\underbrace{\Estat-\gamstat(L_2)}_{\sps
\mbox{$a\rightarrow 0$ in HQET}}
\,+\,\underbrace{\gamstat(L_2)-\gamstat(L_0)}_{\sps
\mbox{$a\rightarrow 0$ in HQET}}
\label{eq2solve}\\
&       &
\,+\,\underbrace{\Gamma(L_0,\Mb)}_{\sps
\mbox{$a\rightarrow 0$ in QCD}}
\,\,\,\,\,\,+\,\,\,\,\,\,\Or\left(\Lambda^2/\Mb\right)
\nonumber
\eea
(with $\Lambda$ a typical low-energy QCD scale), where the terms are just 
arranged such that (the unknown) $\mhbare$ cancels in the energy differences
$\delE\equiv\Estat-\gamstat(L_2)$ and $\gamstat(L_2)-\gamstat(L_0)$,
and the continuum limit exists separately for each of the pieces 
entering \eq{eq2solve}.

The entire heavy quark mass dependence is contained in $\Gamma(L_0,M)$, 
defined in QCD with a relativistic b-quark.
This mass dependence has been non-perturbatively map\-ped out 
in \Ref{QCDvsHQET:pap1}, where as a particular ingredient of the numerical 
calculation, which demands to keep fixed the (dimensionless) RGI heavy quark 
mass while approaching the continuum limit, the knowledge of several 
renormalization factors and improvement coefficients relating the bare to 
the RGI quark mass is required.
Although they had already been determined \cite{msbar:pap1,impr:babp}, it 
was desirable to improve their precision and to estimate them directly in 
the bare coupling range relevant for our application. 
%
%%%%%% figure: CL extrapolations of L_0*Gamma
%
\begin{figure}[htb]
\centering
\vspace{-1.5cm}
\epsfig{file=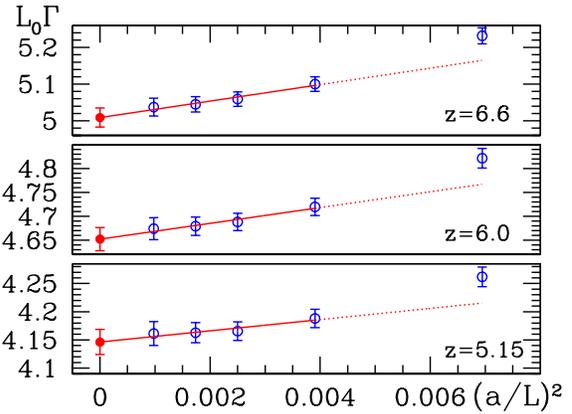,width=8.0cm}
\vspace{-1.375cm}
\caption{
Continuum limit extrapolations of $L_0\Gamma(L_0,M)$ in a relativistic QCD 
and small volume (of linear extent $L_0\approx 0.2\,\Fm$) for representative 
values of $z\equiv L_0M$ \protect\cite{QCDvsHQET:pap1}.
}\label{fig:Gamma}
\vspace{-0.25cm}
\end{figure}
They were thus redetermined in \cite{QCDvsHQET:pap1} and, as exemplified 
in \Fig{fig:Gamma}, performing controlled continuum extrapolations provides 
$\omega\equiv\lim_{a/L\to 0}L_0\Gamma(L_0,M)$ as function of 
$z\equiv L_0M$.
In view of (\ref{eq2solve}) the b-quark mass may now be extracted from the 
interception point of $\omega(z)$ with the combination
\be
\omega_{\rm stat}\equiv
L_0\mB-L_0\left\{\gamstat(L_2)-\gamstat(L_0)\right\}-L_0\delE\,.
\ee

%
%%%%%% figure: graphical solution for M_b
%
\begin{figure}[htb]
\centering
\vspace{-2.25cm}
\epsfig{file=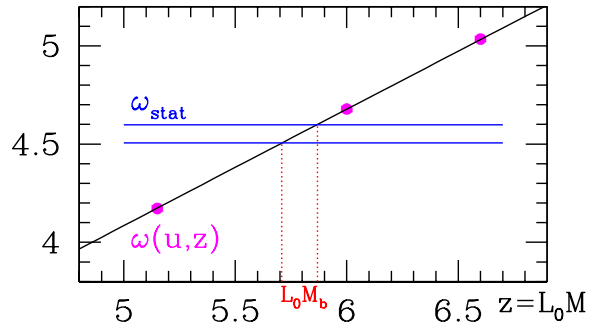,width=8.0cm}
\vspace{-2.0cm}
\caption{
Solution of (\ref{eq2solve}) for the dimensionless RGI b-quark mass, 
$L_0\Mb$.
($u$ indicates that during the computation one works at a certain 
fixed value of the renormalized SF coupling.)
}\label{fig:solveMb}
\vspace{-0.375cm}
\end{figure}
\begin{figure}[htb]
\centering
\vspace{-2.375cm}
\epsfig{file=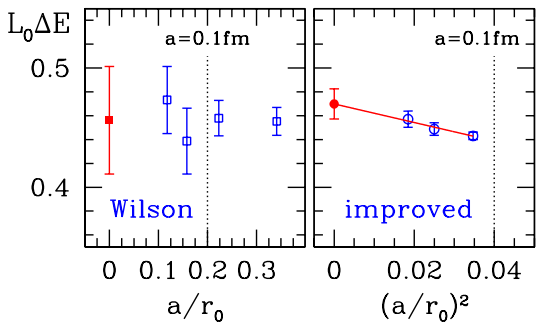,width=8.0cm}
\vspace{-1.75cm}
\caption{
Subtracted, dimensionless $\Bs$--meson energy evaluated from bare Wilson 
fermion numbers of \protect\cite{stat:fnal2} (left) and from simulations 
with the alternative discretization 
of \protect\cite{fbstat:pap1,mbstat:pap2} (right).
}\label{fig:delEstat}
\vspace{-1.0cm}
\end{figure}

The associated graph is given by \Fig{fig:solveMb}, where for the time being 
we restricted the analysis to unimproved Wilson fermion data for $a\Estat$ 
from the literature \cite{stat:fnal2}, resulting in 
$L_0\delE=0.46(5)$ (cf.~the l.h.s.~of \Fig{fig:delEstat}).
We presently obtain in static and quenched 
approximation \cite{mbstat:pap1}
\be
r_0\Mb=16.12(29) \,\,\rightarrow\,\,
\mbbMS\big(\mbbMS\big)=4.12(8)\,\GeV
\label{res_Mb}
\ee
up to corrections of $\Or(\Lambda^2/\Mb)=\Or(\Lambda/(L_0\Mb))$.
From the --- yet preliminary --- 
r.h.s.~of \Fig{fig:delEstat} \cite{fbstat:pap1,mbstat:pap2} one infers that,
once the computation of $a\Estat$ with the static action discussed above 
(which also has linear $\Or(a)$ lattice artifacts removed) is finished, a 
continuum limit of $L_0\delE$ with a (by a factor $\approx3$) smaller error 
is in sight and will substantially enhance the accuracy of the 
result (\ref{res_Mb}).
%
%%%%%% figure: L_0*Delta_E
%

\section{Conclusions and outlook}
\label{Sec_concl}
This status report on actual work of our collaboration makes evident that, 
by virtue of (mainly two) recent advances, non-perturbative calculations 
using the lattice regularized heavy quark effective theory have reached a 
new quality.
One important ingredient is the use of a modified static action which, for 
the first time, enables to compute B-meson lattice correlation functions 
with good statistical precision in the static approximation for 
$x_0>1\,\Fm$.
As demonstrated both for the $\Bs$-meson decay constant and for the b-quark 
mass, this represents a considerable improvement and has a great impact on 
the achievable precision in B-physics computations employing 
HQET.\footnote{
We note in passing that the results reported here agree well with those by 
a different new method that uses extrapolations in the heavy quark mass of 
finite-volume effects in QCD \cite{mb:roma2,fb:roma2c}.
}
The determination of $\mbbMS$ also applies the other promising 
development, a general strategy how to solve renormalization problems in 
HQET entirely non-perturbatively, taking the continuum limit throughout all 
steps involved.

In the quenched approximation, where all the presented results refer to, the 
quoted uncertainties can (and will) be further reduced.
Moreover, in \emph{inter}polating between data obtained in QCD and in the 
static limit, our result for $\Fbs$ is almost independent of any effective 
theory. 

Finally it is worth to emphasize the interesting potential of these methods 
for systematic and straightforward (albeit technically ambitious) 
extensions.
Since it is one of the benefits of the theoretical concepts addressed here 
that describing the b-quark by an \emph{effective theory} circumvents the 
need for prohibitively large lattices (because it completely eliminates the 
mass scale of the b-quark), they will very likely allow to also go beyond 
the static approximation by inclusion of $1/\mb$--corrections as well as to 
incorporate dynamical fermions without major obstacles.
\begin{acknowledgement}
{\it Acknowledgements.}
I am indebted to my colleagues M.~Della Morte, S.~D\"urr, A.~J\"uttner,
H.~Molke, J.~Rolf, A.~Shindler, R.~Sommer and J.~Wennekers for enjoyable
work in our collaboration.
We thank DESY for time on the APEmille computers at Zeuthen.
This project is also supported by the EU under grant HPRN-CT-2000-00145 and 
the DFG in the SFB/TR 09.
\end{acknowledgement}

\bibliography{lattice_ALPHA}
\bibliographystyle{h-elsevierMy}
\end{document}